\documentclass[english]{article}
\usepackage[]{fontenc}
\usepackage[latin9]{inputenc}
\usepackage[letterpaper]{geometry}
\usepackage{float}
\usepackage{amsthm}
\usepackage{amsmath}
\usepackage{amssymb}
\usepackage{graphicx}
\usepackage{esint}
\usepackage[authoryear]{natbib}

\makeatletter

\floatstyle{ruled}
\newfloat{algorithm}{tbp}{loa}
\providecommand{\algorithmname}{Algorithm}
\floatname{algorithm}{\protect\algorithmname}

\numberwithin{equation}{section}
\numberwithin{figure}{section}

\usepackage{soul}
\usepackage{ulem}
\usepackage{pstricks}
\usepackage{algorithmic}
\usepackage{tikz}

\usepackage{subfig}
\usepackage{color}

\newcommand{\deltext}[1]{}

\usepackage{stmaryrd}

\makeatother

\usepackage{babel}
\begin{document}
\begin{center}
{\huge{Spectral analysis of the Navier-Stokes equations using the
combination matrix}}
\par\end{center}{\huge \par}

\begin{center}
\textsc{Lawrence C. Cheung}%
\footnote{\textsc{email: cheung@ge.com}%
}\textsc{ and Tamer Zaki }%
\footnote{\textsc{email: t.zaki@jhu.edu}%
}
\par\end{center}
\begin{quote}
\begin{center}
April 2016
\par\end{center}
\end{quote}

\section{Introduction}

This work is a continuation of the analysis first presented in \citet{CheungZaki2014}.
In that study, the combination matrix was introduced as a means to
tractably handle the nonlinear terms in the spectral domain. An energy
equation was derived from the Navier-Stokes and applied to homongenous
isotropic turbulence, yielding Kolmogorov's -5/3 scaling in the inertial
range.

In this document, a different approach is discussed. Rather than analyze
solutions to the energy equation, we examine the forced Navier-Stokes
equations in spectral space and determine if direct solutions to the
momentum equations can be found. This is done by using the combination
matrix to rewrite the Navier-Stokes as a system of intersecting quadratic
polynomials. Intrepreted geometrically, any solution to the Navier-Stokes
can be represented as a the intersection of a multiple conic sections.
Using the Chebyshev basis, a similar formulation for wall-bounded
channel flow can also be found.

We then find that the tools of commutative algebra can be applied
to determine the solvability of such nonlinear systems. Furthermore,
through the use of polynomial resultants and Groebner bases, all possible
solutions to the systems can be found. This processes is demonstrated
on a simple nonlinear ODE, although it can be extended to more complicated
applications.

\textbf{}

\section{\label{sec:Mathematical-formulation}Mathematical formulation}

\subsection{Homogeneous turbulence}

Here we consider the governing Navier-Stokes equations for the fluid
velocity $\mathbf{u}(\mathbf{x},t)$, pressure $\mathbf{p}(\mathbf{x},t)$,
given a uniform fluid of density $\rho$, viscosity $\nu$, and body
force $\mathbf{f}(\mathbf{x},t)$.

\begin{subequations}\label{NavierStokes}

\begin{equation}
\boldsymbol{\nabla}\cdot\mathbf{u}=0\label{eq:CMass}
\end{equation}

\begin{equation}
\frac{\partial\mathbf{u}}{\partial t}+\mathbf{u}\cdot\boldsymbol{\nabla}\mathbf{u}+\frac{1}{\rho}\nabla p-\nu\nabla^{2}\mathbf{u}+\mathbf{f}=0\label{eq:CMom}
\end{equation}
\end{subequations}For the sake of algebraic simplicity later on,
we assume that the body force $\mathbf{f}$ is divergence free: $\nabla\cdot\mathbf{f}=0$.
However, the analysis can also be repeated for the general case of
arbitrary $\mathbf{f}$ without significant modification. For the
pressure field, velocity fields, and body forces we express them in
terms of a Fourier expansion: \begin{subequations}\label{FourierRep}

\begin{equation}
\mathbf{u}(\mathbf{x},t)=\sum_{\mathbf{k}}\hat{\mathbf{u}}^{\mathbf{k}}e^{ik_{0}t}e^{ik_{1}x_{1}+ik_{2}x_{2}+ik_{3}x_{3}},\label{eq:UFourierRep}
\end{equation}

\begin{equation}
p(\mathbf{x},t)=\sum_{\mathbf{k}}\hat{p}^{\mathbf{k}}e^{ik_{0}t}e^{ik_{1}x_{1}+ik_{2}x_{2}+ik_{3}x_{3}}.\label{eq:PFourierRep}
\end{equation}

\begin{equation}
\mathbf{f}(\mathbf{x},t)=\sum_{\mathbf{k}}\hat{\mathbf{f}}^{\mathbf{k}}e^{ik_{0}t}e^{ik_{1}x_{1}+ik_{2}x_{2}+ik_{3}x_{3}},\label{eq:FFourierRep}
\end{equation}
\end{subequations}The notation used here is similar to the notation
used in \citet{CheungZaki2014}. The superscript on the Fourier coefficient
$\hat{a}_{i}^{\mathbf{k}}$ refers to the wavenumber $\mathbf{k}$,
while the subscript refers to the spatial index $i=1,2,3$ of the
variable. Note, however, that the wavenumber $\mathbf{k}$ in this
context has four components: $\mathbf{k}=\left(k_{0},k_{1},k_{2},k_{3}\right)$,
where $k_{0}$ is the temporal frequency and is synonymous with the
frequency $\omega$. However, please note that we define $k^{2}=k_{1}^{2}+k_{2}^{2}+k_{3}^{2}$,
and summations over the spatial coordinates are taken to be over $i,j=1,2,3$
only. Furthermore, the summation convention is \emph{not} adopted
throughout this analysis in order to avoid any confusion.

To analyze the nonlinear terms, we introduce the combination matrix
$C^{\mathbf{pq},\mathbf{k}}$, which is defined as

\begin{equation}
C^{\mathbf{pq,k}}=\begin{cases}
0, & \mathbf{p}+\mathbf{q}\neq\mathbf{k}\\
1, & \mathbf{p}+\mathbf{q}=\mathbf{k}
\end{cases}.\label{eq:Cpqk_3D_def}
\end{equation}
Using this definition, we can easily represent the multiplication
of two functions $f(\mathbf{x})$ and $g(\mathbf{x})$ in spectral
space. For instance, if $h(\mathbf{x})=f(\mathbf{x})g(\mathbf{x})$
and the functions can be written in terms of the Fourier expansions
\[
h(\mathbf{x})=\sum_{\mathbf{k}}\hat{h}^{\mathbf{k}}e^{i\mathbf{k}\cdot\mathbf{x}}\qquad f(\mathbf{x})=\sum_{\mathbf{p}}\hat{f}^{\mathbf{p}}e^{i\mathbf{p}\cdot\mathbf{x}}\qquad g(\mathbf{x})=\sum_{\mathbf{q}}\hat{g}^{\mathbf{q}}e^{i\mathbf{q}\cdot\mathbf{x}}
\]
then the combination matrix allows us to express the convolution in
terms of a bilinear product, i.e., 

\begin{equation}
\hat{h}^{\mathbf{k}}=\sum_{\mathbf{p}}\sum_{\mathbf{q}}\hat{f}^{\mathbf{p}}C^{\mathbf{pq,k}}\hat{g}^{\mathbf{q}}.\label{eq:hk_mult_expression}
\end{equation}
As discussed in \citet{CheungZaki2014}, this treatment allows the
nonlinear terms of the Navier-Stokes to represented in a tractable
manner. For instance, the nonlinear convective term for the wavenumber
$\mathbf{k}$ can be expressed in spectral space as 
\begin{equation}
\left(\mathbf{u}\cdot\boldsymbol{\nabla}\mathbf{u}\right)^{\mathbf{k}}=i\sum_{\mathbf{p,q}}\left(\begin{array}{c}
\left(\hat{u}_{1}^{\mathbf{p}}q_{1}C^{\mathbf{pq,k}}\hat{u}_{1}^{\mathbf{q}}+\hat{u}_{2}^{\mathbf{p}}q_{2}C^{\mathbf{pq,k}}\hat{u}_{1}^{\mathbf{q}}+\hat{u}_{3}^{\mathbf{p}}q_{3}C^{\mathbf{pq,k}}\hat{u}_{1}^{\mathbf{q}}\right)\\
\left(\hat{u}_{1}^{\mathbf{p}}q_{1}C^{\mathbf{pq,k}}\hat{u}_{2}^{\mathbf{q}}+\hat{u}_{2}^{\mathbf{p}}q_{2}C^{\mathbf{pq,k}}\hat{u}_{2}^{\mathbf{q}}+\hat{u}_{3}^{\mathbf{p}}q_{3}C^{\mathbf{pq,k}}\hat{u}_{2}^{\mathbf{q}}\right)\\
\left(\hat{u}_{1}^{\mathbf{p}}q_{1}C^{\mathbf{pq,k}}\hat{u}_{3}^{\mathbf{q}}+\hat{u}_{2}^{\mathbf{p}}q_{2}C^{\mathbf{pq,k}}\hat{u}_{3}^{\mathbf{q}}+\hat{u}_{3}^{\mathbf{p}}q_{3}C^{\mathbf{pq,k}}\hat{u}_{3}^{\mathbf{q}}\right)
\end{array}\right).\label{eq:Convterm_1}
\end{equation}
If we define the matrix $\boldsymbol{\Gamma}_{ij,m}^{\mathbf{pq},\mathbf{k}}$
as 

\begin{equation}
\boldsymbol{\Gamma}_{ij,m}^{\mathbf{pq},\mathbf{k}}=\left(\begin{array}{ccc}
0 & 0 & q_{1}C^{\mathbf{pq},\mathbf{k}}\\
0 & 0 & q_{2}C^{\mathbf{pq},\mathbf{k}}\\
0 & 0 & q_{3}C^{\mathbf{pq},\mathbf{k}}
\end{array}\right)\left(\begin{array}{ccc}
0 & \mathbf{I} & 0\\
0 & 0 & \mathbf{I}\\
\mathbf{I} & 0 & 0
\end{array}\right)^{(m)}\label{eq:Gamma_pqk_def}
\end{equation}
where the matrix powers $m=1,2,3$ and $\mathbf{I}\equiv\delta^{\mathbf{pq}}$
is the infinite identity matrix, then we can compactly represent the
entire nonlinear convective term as 
\begin{equation}
(\mathbf{u}\cdot\nabla\mathbf{u})_{m}^{\mathbf{k}}=i\sum_{\mathbf{p,q}}\sum_{i,j}\hat{u}_{i}^{\mathbf{p}}\boldsymbol{\Gamma}_{ij,m}^{\mathbf{pq,k}}\hat{u}_{j}^{\mathbf{q}}.\label{eq:ConvTerm_spectral}
\end{equation}
By taking the divergence of (\ref{eq:CMom}) and invoking the continuity
constraint, the pressure Poisson equation $\nabla^{2}p=-\rho\boldsymbol{\nabla\cdot}\left(\mathbf{u}\cdot\boldsymbol{\nabla}\mathbf{u}\right)$
is obtained. Inserting (\ref{eq:UFourierRep}), and using (\ref{eq:Cpqk_3D_def})
in this equation produces the equivalent form in spectral space:

\begin{equation}
\hat{p}^{\mathbf{k}}=\frac{\rho}{k^{2}}\sum_{i,j}\sum_{\mathbf{p},\mathbf{q}}(p_{i}q_{j})\left(\hat{u}_{j}^{\mathbf{p}}C^{\mathbf{pq,k}}\hat{u}_{i}^{\mathbf{q}}\right).\label{eq:Pterm_3}
\end{equation}
\begin{subequations}Inserting (\ref{eq:ConvTerm_spectral}) and (\ref{eq:Pterm_3})
into (\ref{eq:CMom}) yields the conservation of momentum in spectral
form\label{NS_Spectral} 
\begin{equation}
\left(ik_{0}+\nu k^{2}\right)\hat{u}_{m}^{\mathbf{k}}+i\sum_{\mathbf{p,q}}\sum_{i,j}\hat{u}_{i}^{\mathbf{p}}\boldsymbol{\Gamma}_{ij,m}^{\mathbf{pq,k}}\hat{u}_{j}^{\mathbf{q}}-\frac{ik_{m}}{k^{2}}\sum_{i,j}\sum_{\mathbf{p},\mathbf{q}}(p_{i}q_{j})\left(\hat{u}_{j}^{\mathbf{p}}C^{\mathbf{pq,k}}\hat{u}_{i}^{\mathbf{q}}\right)+\hat{f}_{m}^{\mathbf{k}}=0.\label{eq:CMom_Spectral}
\end{equation}
When combined with the conservation of mass in spectral form
\begin{equation}
\sum_{i}k_{i}\hat{u}_{i}^{\mathbf{k}}=0,\label{eq:CMass_Spectral}
\end{equation}
 \end{subequations}equations (\ref{NS_Spectral}) form the governing
Navier-Stokes equations written with an explicit dependence on the
wavenumber $\mathbf{k}$.

Before continuing, we can simplify some of the algebra if we define
the linear operator as 

\[
L_{im}^{\mathbf{p}\mathbf{k}}=\left(ip_{0}+\nu p^{2}\right)\delta_{im}\delta^{\mathbf{pk}}
\]
and the nonlinear operator as

\[
N_{ij,m}^{\mathbf{pq},\mathbf{k}}=i\boldsymbol{\Gamma}_{ij,m}^{\mathbf{pq,k}}-\frac{ik_{m}}{k^{2}}p_{i}q_{j}C^{\mathbf{pq},\mathbf{k}}.
\]
This allows equation (\ref{eq:CMom_Spectral}) to be written compactly
as

\begin{equation}
Z_{m}^{\mathbf{k}}(\hat{\mathbf{u}})=\sum_{i}\sum_{\mathbf{p}}L_{im}^{\mathbf{p}\mathbf{k}}\hat{u}_{i}^{\mathbf{p}}+\sum_{ij}\sum_{\mathbf{p,q}}\hat{u}_{i}^{\mathbf{p}}N_{ij,m}^{\mathbf{pq},\mathbf{k}}\hat{u}_{j}^{\mathbf{q}}+\hat{f}_{m}^{\mathbf{k}}=0.\label{eq:Z_Spectral}
\end{equation}
By adding equation (\ref{eq:Z_Spectral}) together with its transpose

\[
Z_{m}^{\mathbf{k}}+\left(Z_{m}^{\mathbf{k}}\right)^{T}=0
\]
to give

\[
\left(\sum_{i}\sum_{\mathbf{p}}L_{im}^{\mathbf{p}\mathbf{k}}\hat{u}_{i}^{\mathbf{p}}\right)+\left(\sum_{i}\sum_{\mathbf{p}}L_{im}^{\mathbf{p}\mathbf{k}}\hat{u}_{i}^{\mathbf{p}}\right)^{T}+\left(\sum_{ij}\sum_{\mathbf{p,q}}\hat{u}_{i}^{\mathbf{p}}N_{ij,m}^{\mathbf{pq},\mathbf{k}}\hat{u}_{j}^{\mathbf{q}}\right)+\left(\sum_{ij}\sum_{\mathbf{p,q}}\hat{u}_{i}^{\mathbf{p}}N_{ij,m}^{\mathbf{pq},\mathbf{k}}\hat{u}_{j}^{\mathbf{q}}\right)^{T}+\hat{f}_{m}^{\mathbf{k}}+\left(\hat{f}_{m}^{\mathbf{k}}\right)^{T}=0,
\]
we can write an equivalent form of (\ref{eq:Z_Spectral}) as 

\begin{equation}
\left(\sum_{i}\sum_{\mathbf{p}}L_{im}^{\mathbf{p}\mathbf{k}}\hat{u}_{i}^{\mathbf{p}}\right)+\left(\sum_{i}\sum_{\mathbf{p}}L_{im}^{\mathbf{p}\mathbf{k}}\hat{u}_{i}^{\mathbf{p}}\right)^{T}+\left(\sum_{ij}\sum_{\mathbf{p,q}}\hat{u}_{i}^{\mathbf{p}}\mathcal{\mathsf{N}}_{ij,m}^{\mathbf{pq},\mathbf{k}}\hat{u}_{j}^{\mathbf{q}}\right)+\hat{f}_{m}^{\mathbf{k}}+\left(\hat{f}_{m}^{\mathbf{k}}\right)^{T}=0\label{eq:Z_Spectral1}
\end{equation}
or
\begin{equation}
\mathsf{Z}_{m}^{\mathbf{k}}=2\left(\sum_{i}\sum_{\mathbf{p}}L_{im}^{\mathbf{p}\mathbf{k}}\hat{u}_{i}^{\mathbf{p}}\right)+\left(\sum_{ij}\sum_{\mathbf{p,q}}\hat{u}_{i}^{\mathbf{p}}\mathcal{\mathsf{N}}_{ij,m}^{\mathbf{pq},\mathbf{k}}\hat{u}_{j}^{\mathbf{q}}\right)+2\hat{f}_{m}^{\mathbf{k}}=0\label{eq:Z_Spectral2}
\end{equation}
where the nonlinear matrix $\textsf{N}_{ij,m}^{\mathbf{pq},\mathbf{k}}$
is defined as

\begin{eqnarray*}
\textsf{N}_{ij,m}^{\mathbf{pq},\mathbf{k}} & = & N_{ij,m}^{\mathbf{pq},\mathbf{k}}+\left(N_{ij,m}^{\mathbf{pq},\mathbf{k}}\right)^{T}\\
 & = & i\left[\boldsymbol{\Gamma}_{ij,m}^{\mathbf{pq,k}}-\frac{k_{m}}{k^{2}}p_{i}q_{j}C^{\mathbf{pq},\mathbf{k}}\right]+i\left[\boldsymbol{\Gamma}_{ji,m}^{\mathbf{qp,k}}-\frac{k_{m}}{k^{2}}p_{j}q_{i}C^{\mathbf{pq},\mathbf{k}}\right].
\end{eqnarray*}
The astute observer will notice that equation (\ref{eq:Z_Spectral1})
appears as a quadratic equation for $\hat{u}_{i}^{p}$. This can be
expressed in a slightly more insightful form as 

\begin{equation}
\boxed{\sum_{i,j}\sum_{\mathbf{p},\mathbf{q}}\left(\hat{u}_{i}^{\mathbf{p}}+\hat{\zeta}_{i}^{\mathbf{p}}\right)^{T}\left(\textsf{N}_{ij,m}^{\mathbf{pq},\mathbf{k}}\right)\left(\hat{u}_{j}^{\mathbf{q}}+\hat{\zeta}_{j}^{\mathbf{q}}\right)=\hat{R}_{m}^{\mathbf{k}}\quad\textrm{for all }\mathbf{k},\, m}\label{eq:NS_Conic1}
\end{equation}
The definition of the origin point $\hat{\zeta}_{i}^{\mathbf{p}}$
can be found by matching the corresponding term in equation (\ref{eq:Z_Spectral1}):

\[
\sum_{i,j}\sum_{\mathbf{p},\mathbf{q}}\left(\hat{\zeta}_{i}^{\mathbf{p}}\right)^{T}\left(\textsf{N}_{ij,m}^{\mathbf{pq},\mathbf{k}}\right)\hat{u}_{j}^{\mathbf{q}}=\sum_{j}\sum_{\mathbf{q}}L_{jm}^{\mathbf{q}\mathbf{k}}\hat{u}_{j}^{\mathbf{q}}
\]
which suggests that

\begin{equation}
\sum_{i}\sum_{\mathbf{p}}\left(\hat{\zeta}_{i}^{\mathbf{p}}\right)^{T}\left(\textsf{N}_{ij,m}^{\mathbf{pq},\mathbf{k}}\right)=L_{jm}^{\mathbf{q}\mathbf{k}}.\label{eq:NS_Conic_1a}
\end{equation}
The same definition can be found if we match the other term

\[
\sum_{i,j}\sum_{\mathbf{p},\mathbf{q}}\left(\hat{u}_{i}^{\mathbf{p}}\right)^{T}\textsf{N}_{ij,m}^{\mathbf{pq},\mathbf{k}}\hat{\zeta}_{j}^{\mathbf{q}}=\left(\sum_{i}\sum_{\mathbf{p}}L_{im}^{\mathbf{p}\mathbf{k}}\hat{u}_{i}^{\mathbf{p}}\right)^{T}
\]
The definition of $\hat{R}_{m}^{k}$ can be found by order to completing
the square in equation (\ref{eq:NS_Conic1}), yielding

\begin{equation}
\hat{R}_{m}^{\mathbf{k}}=\sum_{i,j}\sum_{\mathbf{p},\mathbf{q}}\left(\hat{\zeta}_{i}^{\mathbf{p}}\right)^{T}\textsf{N}_{ij,m}^{\mathbf{pq},\mathbf{k}}\hat{\zeta}_{j}^{\mathbf{q}}-2\hat{f}_{m}^{\mathbf{k}}\label{eq:NS_Conic1b}
\end{equation}

Note that it is possible to relate the equation at any one wavenumber
$\mathbf{k}$ and direction $m$ to any other $\mathbf{k}'$ and $m'$
through the combination matrix $C^{\mathbf{pq},\mathbf{k}}$. This
is possible through the definition of $\boldsymbol{\Gamma}_{ij,m}^{\mathbf{pq},\mathbf{k}}$
in (\ref{eq:Gamma_pqk_def}), the promotion operator $\boldsymbol{\mathcal{P}}$,
and the relation 
\begin{equation}
\mathbf{C}^{,\mathbf{k}}=\left(\boldsymbol{\mathcal{P}}\right)^{\mathbf{k}}\mathbf{C}^{,0}\label{eq:C_promotion}
\end{equation}
-- see \citet{CheungZaki2014} for details. As a result, for a given
forcing function $f(x)$ and a single equation from (\ref{eq:NS_Conic1}),
one can generate the entire system necessary to solve the Navier-Stokes.

At this point we should pause to reflect on the meaning of equation
(\ref{eq:NS_Conic1}). Equation (\ref{eq:NS_Conic1}) is an exact
restatement of the Navier-Stokes equations (\ref{NavierStokes}) written
in spectral space, assuming that the Fourier representation (\ref{FourierRep})
is valid. No further assumptions have been made. In particular, the
flow is not assumed to be isotropic, as the forcing term $\mathbf{\hat{f}}$
can be different in each spatial direction.

However, by writing the Navier Stokes in the form of (\ref{eq:NS_Conic1}),
we have converted the nonlinear partial differential equation to a
pure system of quadratic polynomials. Interpreted geometrically, the
problem is now akin to finding the intersection points of multiple
conic sections (e.g., intersecting two ellipses or parabolas). The
classification of each of conic section remains is always given by
$\textsf{N}_{ij,m}^{\mathbf{pq},\mathbf{k}}$. However, the location
of each conic section depends on the origin point $\hat{\zeta}_{i}^{\mathbf{k}}$
and radius $R_{m}^{\mathbf{k}}$, which are in turn dependent on the
body force $\mathbf{f}$ and the viscosity $\nu$ present in the problem.

Any solution $\hat{u}_{i}^{\mathbf{k}}$ to this system lies at the
intersection of all polynomials specified by equation (\ref{eq:NS_Conic1}).
While there are infinitely many conics (one for each wavenumber \textbf{$\mathbf{k}$},
and spatial direction $m$), the number of solutions can range from
zero to infinity. Note that as one continuously changes either the
viscosity or the body force, the group of valid solutions may jump
from one set of intersections to another.

Fortunately, finding a common root to a system of polynomial equations
(\ref{eq:Z_Spectral}) or (\ref{eq:NS_Conic1}) is an extensively
studied problem in commutative algebra, and many methods exist for
determining the solvability of such equations, as well as computing
all possible solutions for a finite system. However, before discussing
these methods in section \ref{sec:Groebner-Basis}, we first formulate
the corresponding set of polynomail equations for a planar channel
flow, and demonstrate that the same methodology hold.

\subsection{Channel flow }

\begin{figure}
\begin{centering}
\includegraphics{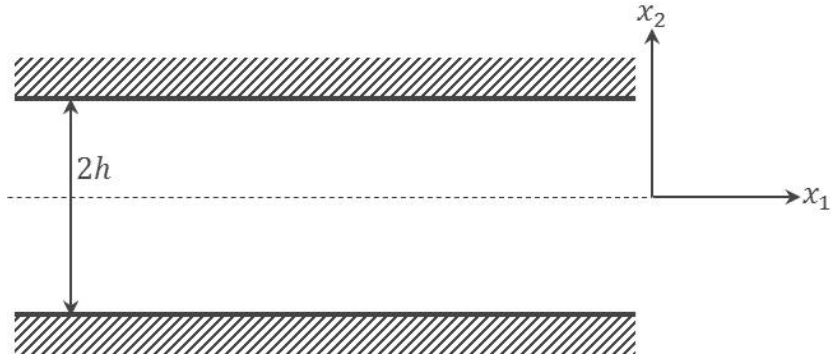}
\par\end{centering}

\caption{\label{fig:Channel-flow}Channel flow}
\end{figure}

The same mathematical formulation can be applied to flow in a different
configuration. In the case of planar channel flow as illustrated in
figure \ref{fig:Channel-flow}, the $x_{1}$ and $x_{3}$ directions
remain homogeneous, but the wall normal direction $x_{2}$ is constrained
by two walls at $x_{2}=\pm h$. For pressure driven channel flow,
we set the applied body force to $\mathbf{f}=\left(\frac{1}{\rho}\overline{\frac{\partial P}{\partial x_{1}}},0,0\right)$.
In this case, we also non-dimensionalize all lengths by the channel
half-height $h$, which yields the following boundary conditions in
the $x_{2}$ direction

\begin{equation}
\mathbf{u}(x_{2}=1)=\mathbf{u}(x_{2}=-1)=0\label{eq:ChannelFlow_BC}
\end{equation}
and corresponding periodic boundaries imposed in the $x_{1}$ and
$x_{3}$ directions.

\subsubsection{\label{sub:Chebyshev-basis}Chebyshev basis}

As an initial foray into the discussion of channel flow, we will adopt
the following Fourier-Chebyshev expansion for the velocity and pressure
fields:

\begin{subequations}\label{FourierChebyshevRep}

\begin{equation}
\mathbf{u}(\mathbf{x},t)=\sum_{\mathbf{k}}\hat{\mathbf{u}}^{\mathbf{k}}e^{ik_{0}t}T_{k_{2}}(x_{2})e^{ik_{1}x_{1}+ik_{3}x_{3}},\label{eq:UFourierChebyRep}
\end{equation}

\begin{equation}
p(\mathbf{x},t)=\sum_{\mathbf{k}}\hat{p}^{\mathbf{k}}e^{ik_{0}t}T_{k_{2}}(x_{2})e^{ik_{1}x_{1}+ik_{3}x_{3}}.\label{eq:PFourierChebyRep}
\end{equation}

\begin{equation}
\mathbf{f}(\mathbf{x},t)=\sum_{\mathbf{k}}\hat{\mathbf{f}}^{\mathbf{k}}e^{ik_{0}t}T_{k_{2}}(x_{2})e^{ik_{1}x_{1}+ik_{3}x_{3}},\label{eq:FFourierChebyRep}
\end{equation}
\end{subequations}where $k_{2}=0,\,1,\,2...\infty$ and $k_{0,}\, k_{1}\,,k_{3}=-\infty,...-1,\,0,\,1,..\infty$.
Although (\ref{FourierChebyshevRep}) does not immediately satisfy
the boundary conditions (\ref{eq:ChannelFlow_BC}), it does allow
us to explore the nonlinear terms without complicating the algebra
at this stage -- a fuller discussion of the boundary conditions appears
in section \ref{sub:Satisfying-boundary-conditions}.

First, however, the Chebyshev polynomials in representation (\ref{FourierChebyshevRep})
require a choice of orthogonalization in the $x_{2}$ direction. Here
we adopt the following inner product between two Chebyshev polynomials
$T_{n}$ and $T_{m}$:

\begin{equation}
\left\langle T_{n}(x),\, T_{m}(x)\right\rangle =\int_{-1}^{1}T_{n}(x)T_{m}(x)\frac{1}{\sqrt{1-x^{2}}}\textrm{d}x=\begin{cases}
0, & n\neq m\\
\pi, & n=m=0\\
\pi/2, & n=m\neq0
\end{cases}\label{eq:Cheby_orthog}
\end{equation}
For Chebyshev polynomials, the equivalent product-to-sum rule is

\begin{equation}
T_{j}(x)T_{k}(x)=\frac{1}{2}\left(T_{j+k}(x)+T_{\left|k-j\right|}(x)\right).\label{eq:Cheby_prod_to_sum}
\end{equation}
Inserting (\ref{eq:Cheby_prod_to_sum}) in (\ref{eq:Cheby_orthog})
and making use of the scalar combination function $C^{pq,k}$ results
in 

\[
\begin{alignedat}{1}\left\langle T_{p}T_{q},\, T_{k}\right\rangle  & =\pi C^{pq,k}\left(\frac{1+\delta^{k0}}{2}\right)+\frac{\pi}{2}\left(C^{p(-q),k}+C^{(-p)q,k}\right)\\
 & \equiv\mathfrak{C}^{pq,k}
\end{alignedat}
\]
and yields the appropriate combination function $\mathfrak{C}^{pq,k}$
for Chebyshev polynomials. To express the multiplication of two functions
$h(\mathbf{x},t)=f(\mathbf{x},t)\cdot g(\mathbf{x},t)$, where each
function is defined as 

\[
f(\mathbf{x},t)=\sum_{\mathbf{p}}\hat{f}^{\mathbf{p}}e^{ip_{0}t}T_{p_{2}}(x_{2})e^{ip_{1}x_{1}+ip_{3}x_{3}},\qquad g(\mathbf{x},t)=\sum_{\mathbf{q}}\hat{g}^{\mathbf{q}}e^{iq_{0}t}T_{q_{2}}(x_{2})e^{iq_{1}x_{1}+iq_{3}x_{3}},
\]
\[
h(\mathbf{x},t)=\sum_{\mathbf{k}}\hat{h}^{\mathbf{k}}e^{ik_{0}t}T_{k_{2}}(x_{2})e^{ik_{1}x_{1}+ik_{3}x_{3}}
\]
the spectral coefficients for $\hat{h}^{\mathbf{k}}$ can be written
as 
\[
\hat{h}^{\mathbf{k}}=\sum_{\mathbf{p}}\sum_{\mathbf{q}}\hat{f}^{\mathbf{p}}C^{p_{0}q_{0},k_{0}}C^{p_{1}q_{1},k_{1}}\mathfrak{C}^{p_{2}q_{2},k_{2}}C^{p_{3}q_{3},k_{3}}\hat{g}^{\mathbf{q}}.
\]
If we define a composite Fourier-Chebyshev combination matrix as
\begin{equation}
\mathsf{\bar{C}}^{\mathbf{pq},\mathbf{k}}=C^{p_{0}q_{0},k_{0}}C^{p_{1}q_{1},k_{1}}\mathfrak{C}^{p_{2}q_{2},k_{2}}C^{p_{3}q_{3},k_{3}}\label{eq:CompositeCpqk}
\end{equation}
then the following analogue to (\ref{eq:hk_mult_expression}) can
be compactly expressed as

\[
\hat{h}^{\mathbf{k}}=\sum_{\mathbf{p}}\sum_{\mathbf{q}}\hat{f}^{\mathbf{p}}\mathsf{\bar{C}}^{\mathbf{pq},\mathbf{k}}g^{\mathbf{q}}.
\]

\subsubsection{\label{sub:ChebyDiffMatrices}Differentiation matrices}

In addition to the multiplication of functions, we also will need
to handle the differentiation of Chebyshev polynominals using this
notation. For the function $f'=df/dy$, we examine the following equalities:

\[
\frac{d}{dy}f(y)=f'(y)=\sum_{k}\hat{f}'^{k}T_{k}(y)=\sum_{k}\hat{f}^{k}\frac{d}{dy}T_{k}(y).
\]
From \citet{peyret2013spectral} the relationship between the coefficients
$\hat{f}'^{k}$ and $\hat{f}^{k}$ can be expressed as 
\[
\hat{f}^{k}=\frac{2}{1+\delta^{k0}}\sum_{\begin{array}{c}
p=k+1\\
(p+k)\,\textrm{odd}
\end{array}}p\hat{f}^{p}
\]
This allows us to express the coefficients $\hat{f}'^{k}$ in terms
of the original coefficients $\hat{f}^{p}$ using a differentiation
matrix $\left(D'_{2}\right)^{kp}$:

\[
\hat{f'}^{k}=\sum_{p}\left(D_{2}^{'}\right)^{kp}\hat{f}^{p},
\]
where the subscript 2 on $D_{2}^{'}$ indicates the derivative with
respect to the $x_{2}$ direction. If we define an equivalent 

\[
\left(D_{1}^{'}\right)^{\mathbf{pq}}=-iq_{1}\delta^{\mathbf{pq}},\quad\left(D_{3}^{'}\right)^{\mathbf{pq}}=-iq_{3}\delta^{\mathbf{pq}},
\]
then the resulting gradient matrix is 

\[
\boldsymbol{\nabla}=\left(\begin{array}{ccc}
\mathbf{D}'_{1}\\
 & \mathbf{D}'_{2}\\
 &  & \mathbf{D}'_{3}
\end{array}\right)=\left(\begin{array}{ccc}
ik_{1}\delta^{\mathbf{pq}}\\
 & \mathbf{D}'_{2}\\
 &  & ik_{3}\delta^{\mathbf{pq}}
\end{array}\right)
\]

A similar process can be used for the second derivaties $f''=d^{2}f/dy^{2}$,
where

\[
\frac{d^{2}}{dy^{2}}f(y)=f''(y)=\sum_{k}\hat{f}''^{k}T_{k}(y)=\sum_{k}\hat{f}^{k}\frac{d^{2}}{dy^{2}}T_{k}(y)
\]
The relationship between the coefficients $\hat{f}^{''k}$ and $\hat{f}^{k}$
can be related using the following formula: 

\[
\hat{f}''^{k}=\frac{1}{1+\delta^{k0}}\sum_{p=k+2}^{\infty}p(p^{2}-k^{2})\left[\frac{1+(-1)^{p+k}}{2}\right]\hat{f}^{p}
\]
(see \citealp{peyret2013spectral}). This relationship can also be
expressed using a second order differentiation matrix $\mathbf{D}_{2}^{''}$
via

\subsubsection*{
\[
\hat{f}''^{k}=\sum_{p}\left(D^{''}\right)^{kp}\hat{f}^{p}.
\]
}

The individual entries in $\mathbf{D}_{2}^{''}$ can be explicitly
written using the following identity for the upper triangular matrix
\[
\mathcal{\mathbf{U}}=\frac{\mathbf{I}}{\mathbf{I}-\boldsymbol{\mathcal{P}}}=\left(\begin{array}{cccc}
1 & 1 & 1\\
 & 1 & 1\\
 &  & 1 & \cdots\\
 &  & \vdots & \ddots
\end{array}\right)
\]
where $\boldsymbol{\mathcal{P}}$ is the promotion operator defined
in \citet{CheungZaki2014}, leading to

\subsubsection*{
\[
\left(\mathbf{D}_{2}^{''}\right)^{kp}=\frac{p(p^{2}-k^{2})}{1+\delta^{k0}}\left[\frac{1+\left(-1\right)^{p+k}}{2}\right]\frac{\left(\boldsymbol{\mathcal{P}}\right)^{2}}{\mathbf{I}-\boldsymbol{\mathcal{P}}}
\]
\textmd{for derivatives in the $x_{2}$ direction. The second order
derivatives in the other directions can be written as} }

\[
\left(D_{1}^{''}\right)^{\mathbf{pq}}=-q_{1}^{2}\delta^{\mathbf{pq}},\quad\left(D_{3}^{''}\right)^{\mathbf{pq}}=-q_{3}^{2}\delta^{\mathbf{pq}}.
\]
From these definitions we can also define a Laplacian operator acting
on the variable $\hat{f}^{\mathbf{q}}$:

\subsubsection*{
\[
\nabla^{2}\hat{f}^{\mathbf{q}}=\boldsymbol{\Delta}^{\mathbf{pq}}\hat{f}^{\mathbf{q}}=\left(\left(D_{1}^{''}\right)^{\mathbf{pq}}+\left(D_{2}^{''}\right)^{\mathbf{pq}}+\left(D_{3}^{''}\right)^{\mathbf{pq}}\right)\hat{f}^{\mathbf{q}},
\]
\textmd{or}}

\[
\boldsymbol{\Delta^{pq}}=\left(\mathbf{D}_{2}^{''}\right)^{\mathbf{pq}}-\left(q_{1}^{2}+q_{3}^{2}\right)\delta^{\mathbf{pq}}.
\]

\subsubsection{The Pressure Poisson Equation}

With the Chebyshev basis and differentiation matrices defined in sections
\ref{sub:Chebyshev-basis} and \ref{sub:ChebyDiffMatrices}, the pressure
Poisson equation

\begin{equation}
\nabla^{2}p=-\rho\boldsymbol{\nabla\cdot}\left(\mathbf{u}\cdot\boldsymbol{\nabla}\mathbf{u}\right)\label{eq:PressPoissonEqn}
\end{equation}
can then be solved for pressure. Assuming the flow is compressible,
the source term on the right hand side is proportional to 

\[
S=\nabla\cdot\mathbf{u}\nabla\mathbf{u}=\sum_{i,j}\frac{\partial u_{i}}{\partial x_{j}}\frac{\partial u_{j}}{\partial x_{i}}.
\]
At each wavenumber $\mathbf{k},$ the source term $S$ can be expressed
as

\begin{equation}
S^{\mathbf{k}}=\left(\begin{array}{c}
\hat{u}_{1}^{\mathbf{p}}\\
\hat{u}_{2}^{\mathbf{p}}\\
\hat{u}_{3}^{\mathbf{p}}
\end{array}\right)\left(\begin{array}{ccc}
\left(-p_{1}\mathsf{\bar{C}}^{,\mathbf{k}}q_{1}\right) & \left(\mathbf{D_{2}^{'}}\right)^{T}\mathsf{\bar{C}}^{,\mathbf{k}}\left(iq_{1}\right) & \left(-p_{3}\mathsf{\bar{C}}^{,\mathbf{k}}q_{1}\right)\\
\left(ip_{1}\right)\mathsf{\bar{C}}^{,\mathbf{k}}\mathbf{D_{2}^{'}} & \left(\mathbf{D_{2}^{'}}\right)^{T}\mathsf{\bar{C}}^{,\mathbf{k}}\mathbf{D_{2}^{'}} & \left(ip_{3}\right)\mathsf{\bar{C}}^{,\mathbf{k}}\mathbf{D_{2}^{'}}\\
\left(-p_{1}\mathsf{\bar{C}}^{,\mathbf{k}}q_{3}\right) & \mathbf{\left(\mathbf{D_{2}^{'}}\right)^{T}}\mathsf{\bar{C}}^{,\mathbf{k}}\left(iq_{3}\right) & \left(-p_{3}\mathsf{\bar{C}}^{,\mathbf{k}}q_{3}\right)
\end{array}\right)\left(\begin{array}{c}
\hat{u}_{1}^{\mathbf{q}}\\
\hat{u}_{2}^{\mathbf{q}}\\
\hat{u}_{3}^{\mathbf{q}}
\end{array}\right)\label{eq:PressPoissonSource}
\end{equation}
using the combination matrix $\overline{\mathsf{C}}^{,\mathbf{k}}$
and differentiation matrices $\mathbf{D}_{2}^{'}$ defined above.
Here we relabel the interior matrix as $B^{\mathbf{pq},\mathbf{k}}$,
which allows (\ref{eq:PressPoissonSource}) to be compactly written
as 

\begin{equation}
S^{\mathbf{k}}=\sum_{\mathbf{p},\mathbf{q}}\hat{\mathbf{u}}^{\mathbf{p}}B^{\mathbf{pq},\mathbf{k}}\hat{\mathbf{u}}^{\mathbf{q}}.\label{eq:PressPoissonEqn2}
\end{equation}
Inserting (\ref{eq:PressPoissonEqn2}) into (\ref{eq:PressPoissonEqn})
and inverting to solve for pressure gives

\begin{equation}
\hat{p}^{\mathbf{k}}=-\rho\sum_{\mathbf{l}}\left(\Delta^{-1}\right)^{\mathbf{kl}}\sum_{\mathbf{p},\mathbf{q}}\hat{\mathbf{u}}^{\mathbf{p}}B^{\mathbf{pq},\mathbf{l}}\hat{\mathbf{u}}^{\mathbf{q}}=-\rho\sum_{\mathbf{p},\mathbf{q}}\sum_{\mathbf{l}}\hat{\mathbf{u}}^{\mathbf{p}}\left(\Delta^{-1}\right)^{\mathbf{kl}}B^{\mathbf{pq},\mathbf{l}}\hat{\mathbf{u}}^{\mathbf{q}}.\label{eq:PressPoissonSource2}
\end{equation}

\subsubsection{The Navier-Stokes equations}

Once the pressure has been found in (\ref{eq:PressPoissonSource2}),
it can be inserted in the Navier-Stokes. From the gradient of pressure

\[
\nabla p=\left(\begin{array}{ccc}
iq_{1}\delta^{\mathbf{pq}} & 0 & 0\\
0 & \left(\mathbf{D}_{2}^{'}\right)^{\mathbf{pq}} & 0\\
0 & 0 & iq_{3}\delta^{\mathbf{pq}}
\end{array}\right)\left(\begin{array}{c}
\hat{p}^{\mathbf{q}}\\
\hat{p}^{\mathbf{q}}\\
\hat{p}^{\mathbf{q}}
\end{array}\right),
\]
each direction $m$ and wavenumber $\mathbf{k}$ component can be
written as

\begin{eqnarray*}
\left(-\frac{1}{\rho}\nabla p\right)_{m}^{\mathbf{k}} & = & \sum_{\mathbf{r}}D_{m}^{\mathbf{\mathbf{kr}}}\sum_{\mathbf{p},\mathbf{q}}\sum_{\mathbf{s}}\hat{\mathbf{u}}^{\mathbf{p}}\left(\Delta^{-1}\right)^{\mathbf{rs}}B^{\mathbf{pq},\mathbf{s}}\hat{\mathbf{u}}^{\mathbf{q}}\\
 & = & \sum_{\mathbf{r},\mathbf{s}}\sum_{\mathbf{p},\mathbf{q}}\hat{\mathbf{u}}^{\mathbf{p}}\left\{ D_{m}^{\mathbf{kr}}\left(\Delta^{-1}\right)^{\mathbf{rs}}B^{\mathbf{pq},\mathbf{s}}\right\} \hat{\mathbf{u}}^{\mathbf{q}}.
\end{eqnarray*}
The corresponding definition of $\boldsymbol{\Gamma}_{ij,m}^{\mathbf{pq},\mathbf{k}}$
for channel flow is

\[
\boldsymbol{\Gamma}_{ij,m}^{\mathbf{pq},\mathbf{k}}=\left(\begin{array}{ccc}
0 & 0 & q_{1}\mathsf{\bar{C}}^{,\mathbf{k}}\\
0 & 0 & -i\mathsf{\bar{C}}^{,\mathbf{k}}\mathbf{D}'_{2}\\
0 & 0 & q_{3}\mathsf{\bar{C}}^{,\mathbf{k}}
\end{array}\right)\left(\begin{array}{ccc}
0 & \mathbf{I} & 0\\
0 & 0 & \mathbf{I}\\
\mathbf{I} & 0 & 0
\end{array}\right)^{(m)}
\]
Combining all of the nonlinear terms in $N_{ij,m}^{\mathbf{pq},\mathbf{k}}$,
we have

\[
N_{ij,m}^{\mathbf{pq},\mathbf{k}}=i\boldsymbol{\Gamma}_{ij,m}^{pq,k}-\sum_{\mathbf{r},\mathbf{s}}\left[D_{m}^{\mathbf{kr}}\left(\Delta^{-1}\right)^{\mathbf{rs}}B^{\mathbf{pq},\mathbf{s}}\right]
\]
Using $\boldsymbol{\Delta}^{\mathbf{pk}}$ defined above, the linear
terms of the Navier-Stokes can be grouped into 

\[
L_{im}^{\mathbf{p}\mathbf{k}}=\left(ip_{0}\delta^{\mathbf{pk}}+\nu\boldsymbol{\Delta}^{\mathbf{pk}}\right)\delta_{im}.
\]
Putting everything together, the Navier-Stokes equations can be written
as

\begin{equation}
Z_{m}^{\mathbf{k}}(\hat{\mathbf{u}})=\sum_{i}\sum_{\mathbf{p}}L_{im}^{\mathbf{p}\mathbf{k}}\hat{u}_{i}^{\mathbf{p}}+\sum_{ij}\sum_{\mathbf{p,q}}\hat{u}_{i}^{\mathbf{p}}N_{ij,m}^{\mathbf{pq},\mathbf{k}}\hat{u}_{j}^{\mathbf{q}}+\hat{f}_{m}^{\mathbf{k}}=0.\label{eq:Z_Spectral_Cheby}
\end{equation}
and it again becomes possible to express the Navier-Stokes equations
as

\[
\sum_{i,j}\sum_{\mathbf{p},\mathbf{q}}\left(\hat{u}_{i}^{\mathbf{p}}+\hat{\zeta}_{i}^{\mathbf{p}}\right)^{T}\left(\textsf{N}_{ij,m}^{\mathbf{pq},\mathbf{k}}\right)\left(\hat{u}_{j}^{\mathbf{q}}+\hat{\zeta}_{j}^{\mathbf{q}}\right)=\hat{R}_{m}^{\mathbf{k}}.
\]

\subsubsection{\label{sub:Satisfying-boundary-conditions}Satisfying boundary conditions}

As mentioned previously, equations (\ref{eq:Z_Spectral_Cheby}) are
not necessarily guaranteeed to satisfy the boundary conditions (\ref{eq:ChannelFlow_BC})
if the Chebyshev polynomials are used directly in representation (\ref{FourierChebyshevRep}).
However, this problem can be solved by taking a special combnation
of Chebyshev functions

\begin{equation}
\phi_{m}(x_{2})=T_{m}(x_{2})-T_{m'}(x_{2})\label{eq:Recombine_Cheby}
\end{equation}
where

\[
m'=\begin{cases}
0, & m\textrm{ even}\\
1, & m\textrm{ odd}
\end{cases}
\]
and the index $m$ now varies from $m=2,3,4,5...$ This ensures that 

\[
\phi_{m}(\pm1)=0
\]
and the no-slip boundary conditions will be automatically satisfied
for any solution of the polynomial system (\ref{eq:Z_Spectral_Cheby}).
Using these functions in the flow field variables gives:\begin{subequations}\label{BasisRecombine}

\begin{equation}
\mathbf{u}(\mathbf{x},t)=\sum_{\mathbf{k}}\hat{\mathbf{u}}^{\mathbf{k}}e^{ik_{0}t}\phi_{k_{2}}(x_{2})e^{ik_{1}x_{1}+ik_{3}x_{3}},\label{eq:UFourierChebyRep-1}
\end{equation}

\begin{equation}
p(\mathbf{x},t)=\sum_{\mathbf{k}}\hat{p}^{\mathbf{k}}e^{ik_{0}t}\phi_{k_{2}}(x_{2})e^{ik_{1}x_{1}+ik_{3}x_{3}}.\label{eq:PFourierChebyRep-1}
\end{equation}

\begin{equation}
\mathbf{f}(\mathbf{x},t)=\sum_{\mathbf{k}}\hat{\mathbf{f}}^{\mathbf{k}}e^{ik_{0}t}\phi_{k_{2}}(x_{2})e^{ik_{1}x_{1}+ik_{3}x_{3}}.\label{eq:FFourierChebyRep-1}
\end{equation}
\end{subequations}It can be shown that the differentiation matrices
$\mathbf{D}'_{2}$ and $\mathbf{D}''_{2}$ do not require modification
with the new basis. However, the combination matrix takes a more complicated
form. If we project the nonlinear product $\phi_{m}\phi_{n}$ back
onto the basis $\phi_{k}$ using the same inner product, we have

\begin{eqnarray*}
\left\langle \phi_{m}\phi_{n},\phi_{k}\right\rangle  & = & \left\langle \left(T_{m}-T_{m'}\right)\left(T_{n}-T_{n'}\right),\left(T_{k}-T_{k'}\right)\right\rangle \\
 & = & \left\langle T_{m}T_{n}-T_{m'}T_{n}-T_{m}T_{n'}+T_{m'}T_{n'},\left(T_{k}-T_{k'}\right)\right\rangle \\
 & = & \left\langle T_{m}T_{n},T_{k}\right\rangle -\left\langle T_{m'}T_{n},T_{k}\right\rangle -\left\langle T_{m}T_{n'},T_{k}\right\rangle +\left\langle T_{m'}T_{n'},T_{k}\right\rangle \\
 &  & +\left\langle T_{m}T_{n},T_{k'}\right\rangle -\left\langle T_{m'}T_{n},T_{k'}\right\rangle -\left\langle T_{m}T_{n'},T_{k'}\right\rangle +\left\langle T_{m'}T_{n'},T_{k'}\right\rangle \\
 & = & \mathfrak{C}^{mn,k}-\mathfrak{C}^{m'n,k}-\mathfrak{C}^{mn',k}+\mathfrak{C}^{m'n',k}+\mathfrak{C}^{mn,k'}-\mathfrak{C}^{m'n,k'}-\mathfrak{C}^{mn',k'}+\mathfrak{C}^{m'n',k'}\\
 & = & \breve{C}^{mn,k}
\end{eqnarray*}
The composite combination function is then
\begin{equation}
\mathsf{\bar{C}}^{\mathbf{pq},\mathbf{k}}=C^{p_{0}q_{0},k_{0}}C^{p_{1}q_{1},k_{1}}\breve{C}^{p_{2}q_{2},k_{2}}C^{p_{3}q_{3},k_{3}}.\label{eq:CompositeCpqk_Cheby2}
\end{equation}
Using the modified composite combination function (\ref{eq:CompositeCpqk_Cheby2})
in the governing equations (\ref{eq:Z_Spectral_Cheby}) then allows
the flow field to satisfy the no-slip boundary conditions (\ref{eq:ChannelFlow_BC}).

\section{\label{sec:Solvability}Solvability }

The results of section (\ref{sec:Mathematical-formulation}) showed
how the nonlinear Navier-Stokes equations could be converted into
a series of quadratic polynomials, whose solution lies at the intersection
of the entire system. Fortunately, a significant amount of literature
in commutative algebra is devoted towards finding the common root
of for systems of polynomials \citep{Emiris:2010:ANA:1882757.1882774,cox1992ideals},
and can be applied in this case. In the next two sections, we explore
how the theory of polynomial resultants and Groebner Basis can be
used to determine the solvability and find all possible spectral solutions
to the Navier-Stokes equations, if they exist.

The solvability of a system of polynomial equations such as (\ref{eq:Z_Spectral2})
or (\ref{eq:Z_Spectral_Cheby}) can be determined through the use
of polynomial resultants. Due to its extensive body of literature,
a complete summary the theory of resultants is omitted here for conciseness,
although additional details can be found in \citet{cox2006using}.
For the purposes of this work, we define the resultant of a system
of polynomials as an algebraic function of their coefficients which
is zero if and only if the polynomials contain a common root. For
example, given two polynomials $f(x)$ and $g(x)$ defined as 

\[
f(x)=a_{n}x^{n}+...+a_{1}x^{1}+a_{0},
\]

\[
g(x)=b_{m}x^{m}+...+b_{1}x^{1}+b_{0}
\]
the resultant $R(f,g)$ of $f$ and $g$ is defined as 

\[
R(f,g)=a_{n}^{m}b_{m}^{n}\left(\prod_{i=1}^{n}\prod_{j=1}^{m}\left(\alpha_{i}-\beta_{j}\right)\right),
\]
where $\alpha_{i}$ and $\beta_{i}$ are the roots of $f$ and $g$,
respectively. In the case where both $f$ and $g$ are second order
polynomials,

\[
f(x)=a_{2}x^{2}+a_{1}x^{1}+a_{0}
\]

\[
g(x)=b_{2}x^{2}+b_{1}x^{1}+b_{0}
\]
the resultant can be defined in terms of the determinant of the Sylvester
matrix \citep{sylvester1853theory} and the coefficients $a_{n}$
and $b_{n}$, or

\[
R(f,g)=\left|\begin{array}{cccc}
a_{2} & a_{1} & a_{0} & 0\\
0 & a_{2} & a_{1} & a_{0}\\
b_{2} & b_{1} & b_{0} & 0\\
0 & b_{2} & b_{1} & b_{0}
\end{array}\right|.
\]
Therefore, if $f$ and $g$ contain a common root, then 

\begin{equation}
R(f,g)=(a_{2}b_{0}-b_{2}a_{0})^{2}-(a_{2}b_{1}-b_{2}a_{1})(a_{1}b_{0}-b_{1}a_{0})=0\label{eq:SylRes_quad1}
\end{equation}

In the case where $f$ and $g$ are functions of multiple independent
variables, the Sylvester resultant is still valid. For example, if
both $f=f(x,y,z)$ and $g=g(x,y,z)$ and are quadratic in $z$, we
can express them as
\[
f(x,y,z)=a_{2}(x,y)z^{2}+a_{1}(x,y)z^{1}+a_{0}(x,y)
\]

\[
g(x,y,z)=b_{2}(x,y)z^{2}+b_{1}(x,y)z^{1}+b_{0}(x,y)
\]
where the coefficients $a_{i}$ and $b_{i}$ are now dependent on
$x$ and $y$. The Sylvester resultant can then be computed with respect
to $z$ such that

\[
R_{z}(f,g)=\left|\begin{array}{cccc}
a_{2}(x,y) & a_{1}(x,y) & a_{0}(x,y) & 0\\
0 & a_{2}(x,y) & a_{1}(x,y) & a_{0}(x,y)\\
b_{2}(x,y) & b_{1}(x,y) & b_{0}(x,y) & 0\\
0 & b_{2}(x,y) & b_{1}(x,y) & b_{0}(x,y)
\end{array}\right|=0
\]

For a system of equations containing more than two polynomials, the
resultant of any two arbitrarily chosen polynomials must also be zero
for there to be a common root for the entire system. This provides
a relatively simple test to determine the solvability of a system
of polynomial equations such as (\ref{eq:Z_Spectral2}) or (\ref{eq:Z_Spectral_Cheby}).
However, before applying the Sylvester resultant to the current problem,
we can make use of a variable substitution to transform the equations
into a more suitable form. If we insert 

\[
\hat{v}_{i}^{\mathbf{k}}=\hat{u}_{i}^{\mathbf{k}}/w
\]
into equation (\ref{eq:Z_Spectral2}), the result can be written as

\begin{equation}
\mathsf{Z}_{m}^{\mathbf{k}}(\hat{\mathbf{v}}^{\mathbf{k}},w)=2\left(\sum_{i}\sum_{\mathbf{p}}L_{im}^{\mathbf{p}\mathbf{k}}\hat{v}_{i}^{\mathbf{p}}w\right)+\left(\sum_{ij}\sum_{\mathbf{p,q}}\hat{v}_{i}^{\mathbf{p}}\mathcal{\mathsf{N}}_{ij,m}^{\mathbf{pq},\mathbf{k}}\hat{v}_{j}^{\mathbf{q}}\right)+2\hat{f}_{m}^{\mathbf{k}}w^{2}=0.\label{eq:Z_Spectral3}
\end{equation}
Here the role of $w$ is to serve as a \emph{homogenizing} variable,
and removes some of the arbitrariness in selecting the variable to
compute the resultant. Note that each term in (\ref{eq:Z_Spectral3})
now has the same degree in either $\hat{v}_{i}^{\mathbf{k}}$ or $w$,
making the polynomial $\mathsf{Z}_{m}^{\mathbf{k}}$ homogeneous (not
to be confused with the homogeneous nature of the flow in the $x_{1}$
and $x_{3}$ directions). We can then compute the resultant of any
two polynomials $\mathsf{Z}_{m}^{\mathbf{k}}$ and $\mathsf{Z}_{m'}^{\mathbf{k'}}$
with respect to $w$

\[
R_{w}(\mathsf{Z}_{m}^{\mathbf{k}},\,\mathsf{Z}_{m'}^{\mathbf{k}'})=\left|\begin{array}{cccc}
2\hat{f}_{m}^{\mathbf{k}}\quad & 2\sum L_{im}^{\mathbf{p}\mathbf{k}}\hat{v}_{i}^{\mathbf{p}}w\quad & \sum_{ij}\sum_{\mathbf{p,q}}\hat{v}_{i}^{\mathbf{p}}\mathcal{\mathsf{N}}_{ij,m}^{\mathbf{pq},\mathbf{k}}\hat{v}_{j}^{\mathbf{q}} & 0\\
0 & 2\hat{f}_{m}^{\mathbf{k}} & 2\sum L_{im}^{\mathbf{p}\mathbf{k}}\hat{v}_{i}^{\mathbf{p}}w & \sum_{ij}\sum_{\mathbf{p,q}}\hat{v}_{i}^{\mathbf{p}}\mathcal{\mathsf{N}}_{ij,m}^{\mathbf{pq},\mathbf{k}}\hat{v}_{j}^{\mathbf{q}}\\
2\hat{f}_{m'}^{\mathbf{k}'} & 2\sum L_{im'}^{\mathbf{p}\mathbf{k'}}\hat{v}_{i}^{\mathbf{p}}w & \sum_{ij}\sum_{\mathbf{p,q}}\hat{v}_{i}^{\mathbf{p}}\mathcal{\mathsf{N}}_{ij,m'}^{\mathbf{pq},\mathbf{k}'}\hat{v}_{j}^{\mathbf{q}} & 0\\
0 & 2\hat{f}_{m'}^{\mathbf{k}'} & 2\sum L_{im'}^{\mathbf{p}\mathbf{k'}}\hat{v}_{i}^{\mathbf{p}}w & \sum_{ij}\sum_{\mathbf{p,q}}\hat{v}_{i}^{\mathbf{p}}\mathcal{\mathsf{N}}_{ij,m'}^{\mathbf{pq},\mathbf{k}'}\hat{v}_{j}^{\mathbf{q}}
\end{array}\right|
\]
After some significant algebraic manipulation, the Sylvester resultant
can be written as

\begin{eqnarray}
R_{w}(\mathsf{Z}_{m}^{\mathbf{k}},\,\mathsf{Z}_{m'}^{\mathbf{k}'}) & =\sum_{\mathbf{p},\mathbf{q}}\sum_{\mathbf{r},\mathbf{s}}\sum_{i,j}\sum_{l,n}\hat{v}_{i}^{\mathbf{p}}\hat{v}_{j}^{\mathbf{q}}\hat{v}_{l}^{\mathbf{r}}\hat{v}_{s}^{\mathbf{n}} & \times\left\{ 4N_{ij,m}^{\mathbf{pq},\mathbf{k}}N_{ln,m}^{\mathbf{rs},\mathbf{k}}\left(\hat{F}_{m'}^{k'}\right)^{2}+4N_{ij,m'}^{\mathbf{pq},\mathbf{k}'}N_{ln,m'}^{\mathbf{rs},\mathbf{k}'}\left(\hat{F}_{m}^{k}\right)^{2}\right.\nonumber \\
 &  & -4N_{ij,m'}^{\mathbf{pq},\mathbf{k}'}N_{ln,m}^{\mathbf{rs},\mathbf{k}}\hat{F}_{m}^{k}\hat{F}_{m'}^{k'}-4N_{ij,m}^{\mathbf{pq},\mathbf{k}}N_{ln,m'}^{\mathbf{rs},\mathbf{k}'}\hat{F}_{m'}^{k'}\hat{F}_{m}^{k}\nonumber \\
 &  & -8L_{im}^{\mathbf{pk}'}\hat{F}_{m}^{\mathbf{k}}L_{lm}^{\mathbf{rk}}N_{ij,m'}^{\mathbf{pq},\mathbf{k}'}-8L_{im}^{\mathbf{pk}}\hat{F}_{m'}^{\mathbf{k}'}L_{lm'}^{\mathbf{rk}'}N_{ij,m}^{\mathbf{pq},\mathbf{k}}\nonumber \\
 &  & \left.+8L_{im'}^{\mathbf{pk}'}\hat{F}_{m}^{\mathbf{k}}L_{lm}^{\mathbf{rk}'}N_{ij,m}^{\mathbf{pq},\mathbf{k}}-8L_{lm}^{\mathbf{rk}}\hat{F}_{m'}^{\mathbf{k}'}L_{im}^{\mathbf{pk}}N_{ij,m'}^{\mathbf{pq},\mathbf{k}'}\right\} =0\label{eq:SylRes_NS1}\\
\nonumber 
\end{eqnarray}
For a common root to exist for the system (\ref{eq:Z_Spectral3}),
equation (\ref{eq:SylRes_NS1}) must be satisified for any choice
of wavenumbers and directions $(\mathbf{k},m)$ and $(\mathbf{k}',m')$.
Note that such a test merely provides a \emph{necessary} condition
for a solution to exist, and not a \emph{sufficient} condition. To
determine the actual solution(s) of the original system, we must apply
some additional concepts from commutative algebra.

\section{\label{sec:Groebner-Basis}Groebner Basis and a model problem}

The results of the previous section provided a set of conditions for
which solutions to (\ref{eq:Z_Spectral3}) exist. However, they did
not provide a means to determine the actual solution(s), if any exist.
To illustrate how this might be done, we introduce the idea of a Groebner
basis and demonstrate its application on a model nonlinear problem.

Before proceeding, we should note that multiple methods exist for
solving systems such as (\ref{eq:Z_Spectral3}), and each method possesses
its own particular set of advantages and disadvantages. The method
of Groebner basis described below is merely one which is relatively
quick to implement for simple problems and provides a solid theoretical
foundation for further exploration. In the section below, we also
consider solutions for a finite set of polynomials.

Defined mathematically, given a set of polynomials $I=\{f_{1}(x_{1},x_{2},...),\, f_{2}(x_{1},x_{2},...),\,...,f_{n}(x_{1},x_{2},...)\}$
and a particular monomial ordering for $(x_{1},x_{2},...)$, the polynomials
$G=\{g_{1},\, g_{2},..,g_{n}\}$ is a Groebner basis of polynomials
if and only if the leading term of any element in $I$ is divisible
by one of the leading terms in $G$. From this definition a significant
number of results can be derived, many of which are discussed in \citep{cox1992ideals,cox2006using}
and other literature. For the current work, though, we focus on two
properties of the Groebner basis which are relevant to our application.
First, it can be shown that for the system of equations \begin{subequations}\label{Ideal_system_eqn}

\begin{eqnarray}
f_{1}(x_{1},x_{2},...) & = & 0\label{eq:}\\
f_{2}(x_{1},x_{2},...) & = & 0\label{eq:-1}\\
 & \vdots\nonumber \\
f_{n}(x_{1},x_{2},...) & = & 0\label{eq:-2}
\end{eqnarray}
\end{subequations}to have a common solution, the Groebner basis cannot
contain the polynomial $g_{i}=1$. This provides a convenient test
for the solvability of a system of polynomial equations.

The second remarkable property is that the set of polynomials in a
Gröbner basis (set equal to zero) has the same collection of roots
as the original polynomials. That is, the solution to 

\begin{subequations}\label{Groebner_system_eqn}

\begin{eqnarray}
g_{1}(x_{1},x_{2},...) & = & 0\label{eq:-3}\\
g_{2}(x_{1},x_{2},...) & = & 0\label{eq:-4}\\
 & \vdots\nonumber \\
g_{n}(x_{1},x_{2},...) & = & 0\label{eq:-5}
\end{eqnarray}
\end{subequations}is the same solution to (\ref{Ideal_system_eqn}).
Furthermore, the system (\ref{Groebner_system_eqn}) can be solved
in a much easier fashion compared to (\ref{Ideal_system_eqn}). Due
to the eliminiation theorem \citep{cox1992ideals}, we are guaranteed
that one of the equations (\ref{Groebner_system_eqn}) will contain
one of the variables $x_{i}$ in isolation, and can be solved for
algebraically. After $x_{i}$ is found, it can be used to eliminate
another variable $x_{i'}$, and so on until the entire system is solution.
In this manner, using the Groebner basis is similar to solving a linear
system of equations by LU decomposition and backsubstiution. In fact,
finding a Groebner basis for a linear system of multiple variaible
is equivalent to Gaussian elimination, and for univariate polynomials,
it is equivalent to the Euclidian algorithm. An added feature to solving
system of polynomial equations via Groebner bases is that all solutions,
if they exist, can be found. 

The first algorithm for generating a Groebner basis from a set of
polynomials was given by Buchberger \citep{buchberger1976theoretical}
and briefly described in Appendix \ref{sec:Appendix}. Since that
seminal work, alternative algorithms have been developed to generate
Groebner basis which are more computationally efficient, but for the
purposes of the current analysis and the model problem discussed below,
Buchberger's simple algorithm more than suffices.

\subsubsection{Model problem}

To illustrate how the Groebner basis can be used to solve nonlinear
differential equations, we consisder a one-dimensional ordinary differential
equation 

\begin{equation}
\frac{d^{2}u}{dx^{2}}+u^{2}-f=0\label{eq:ModProb1}
\end{equation}
for $u(x)$ and a forcing function $f(x)$ over the domain $0\le x\le2\pi$,
with the periodic boundary conditions

\[
u(0)=u(2\pi),\qquad f(0)=f(2\pi).
\]
While not as complex as the Navier-Stokes (\ref{NavierStokes}), equation
(\ref{eq:ModProb1}) contains the necessary nonlinear features to
demonstrate the procedure. Similar to \ref{FourierRep}, we use the
Fourier representation

\begin{equation}
u(x)=\sum_{k}\hat{u}^{k}e^{ikx},\qquad f(x)=\sum_{k}\hat{f}^{k}e^{ikx}\label{eq:Model_fourier_rep}
\end{equation}
Inserting (\ref{eq:Model_fourier_rep}) into (\ref{eq:ModProb1})
and using the combination matrix gives

\begin{equation}
Z^{k}(\hat{u}^{i})=-k^{2}\hat{u}^{k}+\sum_{p,q}\hat{u}^{p}C^{pq,k}\hat{u}^{q}+\hat{f}^{k}=0,\quad\textrm{for }k=-\infty,...-1,0,1,...\infty\label{eq:ModProb2}
\end{equation}
An equivalent representation of (\ref{eq:ModProb2}) for each polynomial
$Z^{k}$ is

\begin{equation}
\hat{Z}^{k}(\hat{u}^{i})=\sum_{p,q}\left(\begin{array}{c}
\hat{u}^{p}\\
1
\end{array}\right)\cdot\left(\begin{array}{cc}
2C^{pq,k} & -p^{2}\delta^{pk}\\
-q^{2}\delta^{qk} & 2\hat{f}^{k}
\end{array}\right)\left(\begin{array}{c}
\hat{u}^{q}\\
1
\end{array}\right),\label{eq:ModProb3}
\end{equation}
However, to apply Buchberger's algorithm, we must choose a finite
set of polynmials. Therefore, we truncate all wavenumbers greater
than $N$ to result in the following system of polynomial equations:

\begin{equation}
Z^{k}(\hat{u}^{i})=-k^{2}\hat{u}^{k}+\sum_{\left|p\right|,\left|q\right|\le N}\hat{u}^{p}C^{pq,k}\hat{u}^{q}+\hat{f}^{k}=0,\quad\textrm{for }k=-N,...-1,0,1,...N\label{eq:ModProb4}
\end{equation}
To apply Buchberger's algorithm to (\ref{eq:ModProb4}), we choose
to use a graded reverse lexographic ordering of monomials. Note that
the choice of ordering does not affect the resulting Groebner basis,
although the efficiency of the algorithm can vary.

\begin{figure}
\begin{centering}
\includegraphics[width=4.25in]{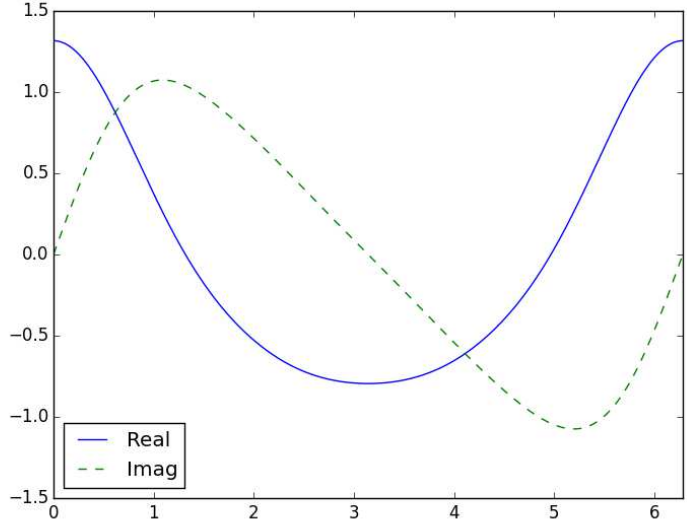}
\par\end{centering}

\caption{\label{fig:Model-solution}The solution to the model problem (\ref{eq:ModProb1})
with $f(x)=e^{ix}$ using $N=4$.}
\end{figure}

In the following example, we choose the forcing function $f(x)=e^{ix}$
and truncate the system of polynomials. For the case where $N=2$,
the Groebner basis polynomials are

\begin{subequations}
\begin{equation} \hat{u}^{-2} = 0 \end{equation}
\begin{equation} \hat{u}^{-1} = 0 \end{equation}
\begin{equation} 9\hat{u}^{0} - 8\hat{u}^{2} + 2 = 0 \end{equation}
\begin{equation} 9\hat{u}^{1} - 16\hat{u}^{2} - 5 = 0                      \end{equation}
\begin{equation} 16(\hat{u}^{2})^2 - 8\hat{u}^{2} + 1 = 0 \label{eq:N2_k2} \end{equation}
\end{subequations}

Note that equation (\ref{eq:N2_k2}) can be solved directly for $\hat{u}^{2}$
and substituted in the remaining equations to generate the one possible
solution to the entire system, yielding

\begin{equation}
\hat{u}^{-2}=\hat{u}^{-1}=\hat{u}^{0}=0,\quad\textrm{and }\hat{u}^{1}=1,\,\hat{u}^{2}=1/4\label{eq:N2_solution}
\end{equation}
If we retain all Fourier coefficients $\hat{u}^{k}$ for $\left|k\right|\le3$,
then the Groebner basis polynomials using graded reverse lexographic
ordering are

\begin{subequations}

\begin{equation} \hat{u}^{-3} = 0 \end{equation} 
\begin{equation} \hat{u}^{-2} = 0 \end{equation}
\begin{equation} \hat{u}^{-1} = 0 \end{equation} 
\begin{equation} 121\hat{u}^{0} - 324\hat{u}^{3} + 18 = 0 \end{equation} 
\begin{equation} 121\hat{u}^{1} - 648\hat{u}^{3} - 85 = 0 \end{equation}
\begin{equation} 242\hat{u}^{2} - 729\hat{u}^{3} - 20 = 0 \end{equation}
\begin{equation} 324(\hat{u}^{3})^2 - 36\hat{u}^{3} + 1 = 0 \end{equation}

\end{subequations}which can be further simplified to give the one possible solution:
\begin{equation}
\hat{u}^{-3}=\hat{u}^{-2}=\hat{u}^{-1}=\hat{u}^{0}=0,\quad\textrm{and }\hat{u}^{1}=1,\,\hat{u}^{2}=1/4,\,\hat{u}^{3}=1/18\label{eq:N3_solution}
\end{equation}

Both solutions given by (\ref{eq:N2_solution}) and (\ref{eq:N3_solution})
can be verified by direct substitution into the original equation
(\ref{eq:ModProb1}). Furthermore, from solutions (\ref{eq:N2_solution})
or (\ref{eq:N3_solution}), one can calculate the remaining coefficients
$\hat{u}^{k}$ for $\left|k\right|>N$ through relatively simple algebra.
A graph of the solution $u(x)$ is shown in figure \ref{fig:Model-solution}.

\begin{figure}
\newlength{\rod}
\setlength{\rod}{2.5cm}
\psset{unit=1\rod}
\begin{pspicture}(7.5,3)
\rput[bl](-.25,0.0){\includegraphics[width=3.95\rod]{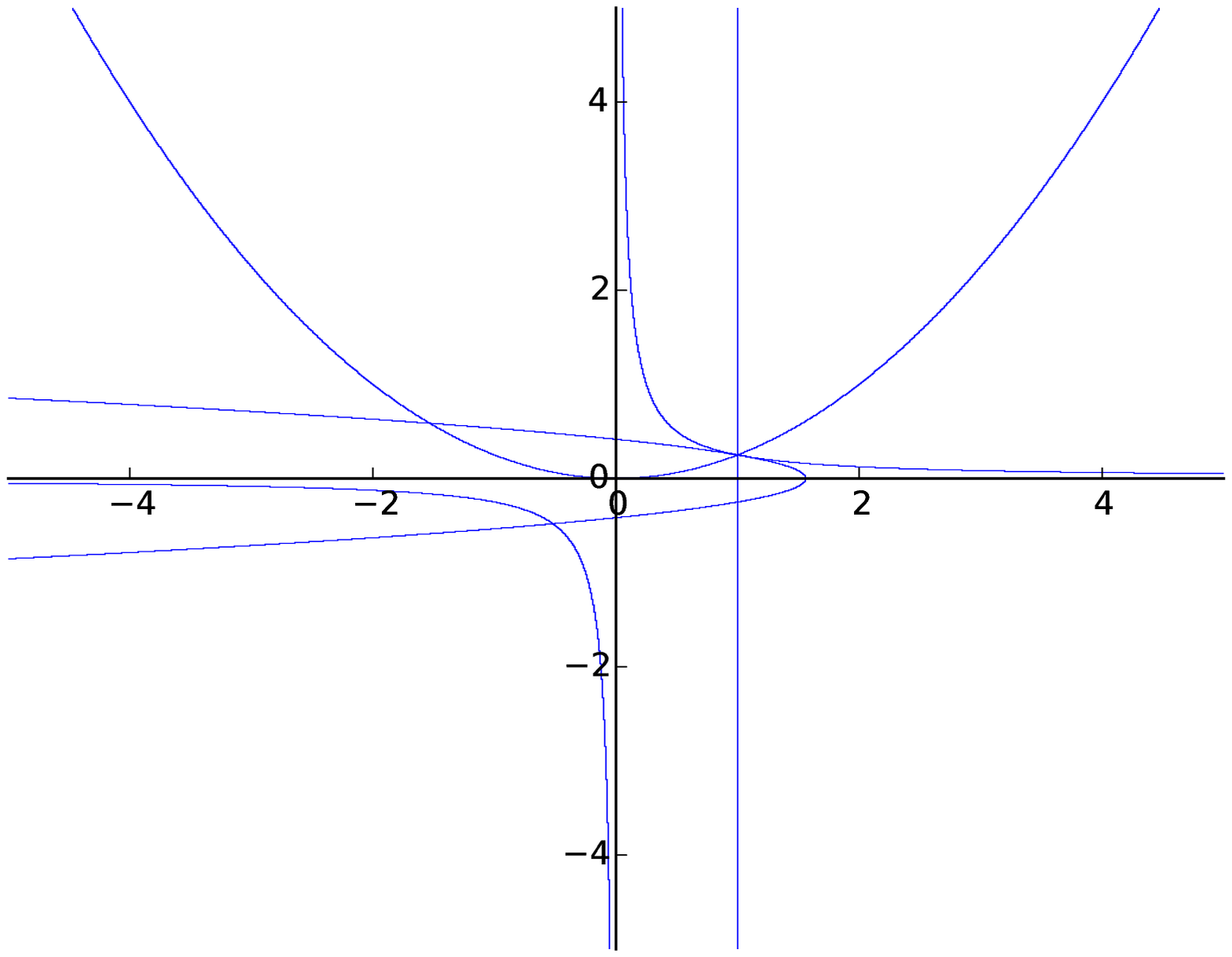}}
\rput[bl](0.20,0.20){(a)}
\rput[bl](3.10,1.30){$\hat{u}^1$}
\rput[bl](1.60,2.60){$\hat{u}^2$}

\rput[bl](2.10,2.20){$Z^1$}
\rput[bl](3.10,2.40){$Z^2$}
\rput[bl](1.85,1.80){$Z^3$}
\rput[bl](0.40,1.70){$Z^4$}

\rput[bl](3.4,0.0){\includegraphics[width=3.95\rod]{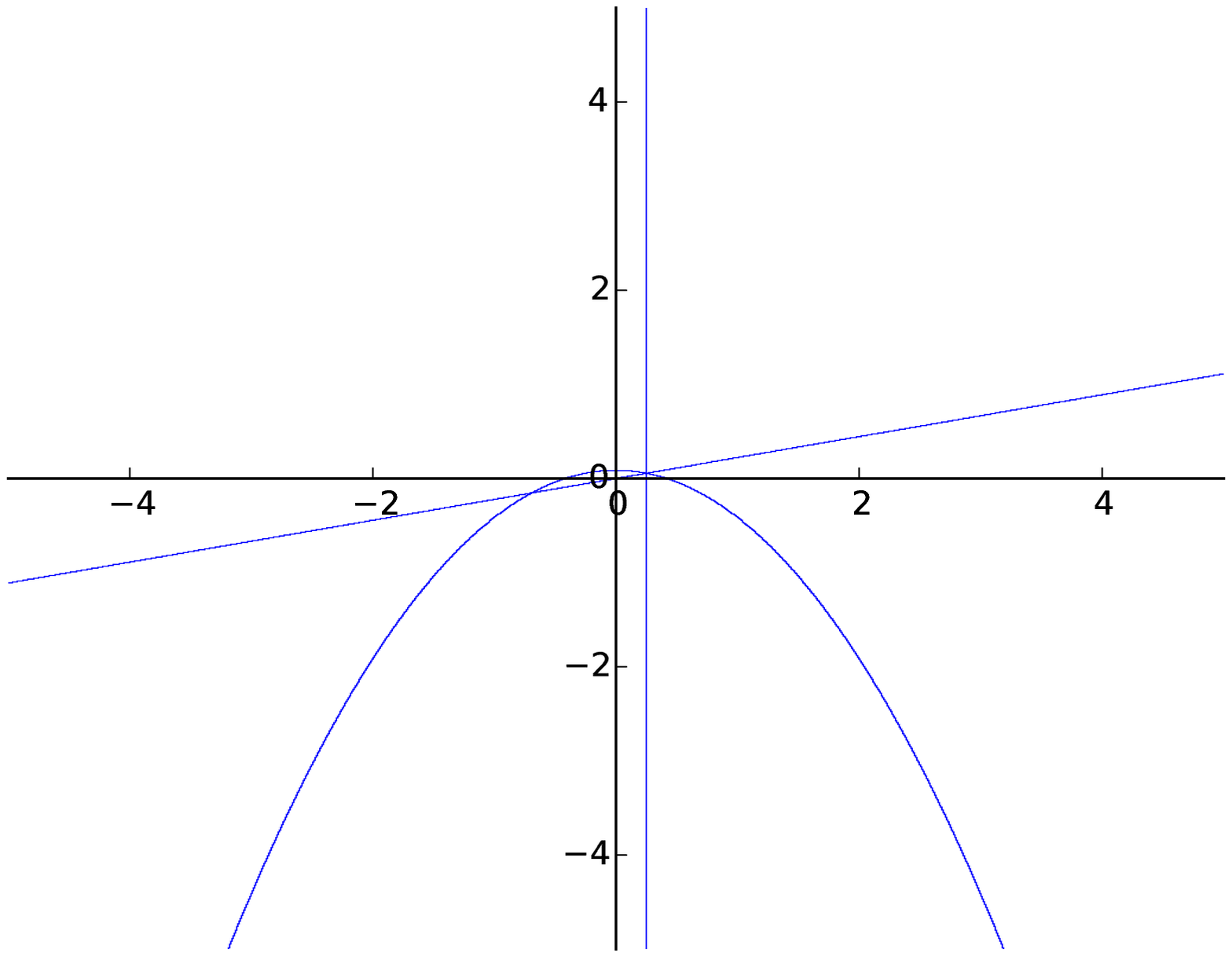}}
\rput[bl](3.80,0.20){(b)}
\rput[bl](6.80,1.30){$\hat{u}^2$}
\rput[bl](5.20,2.60){$\hat{u}^3$}

\rput[bl](5.55,2.10){$Z^2$}
\rput[bl](6.80,1.80){$Z^3$}
\rput[bl](6.10,1.10){$Z^4$}

\end{pspicture}

\caption{\label{fig:PolynomialPlot}Plot of the polynomials (a) $Z^{1}=0$,
$Z^{2}=0$, $Z^{3}=0$,$Z^{4}=0$ in the $\hat{u}^{1}$-$\hat{u}^{2}$
plane, and (b) $Z^{2}=0$, $Z^{3}=0$, $Z^{4}=0$ in the $\hat{u}^{2}$-$\hat{u}^{3}$
plane. The solution lies at the intersection of the displayed curves.}
\end{figure}

A simple visualization of the quadratic polynomials $Z^{k}=0$ is
possible if we allow two coefficients $\hat{u}^{k}$ and $\hat{u}^{k'}$to
vary while fixing all other coefficients at the common root. An example
is shown in figure \ref{fig:PolynomialPlot} where selected polynomials
$Z^{k}(\hat{u}^{1},\hat{u}^{2})=0$ and $Z^{k}(\hat{u}^{2},\hat{u}^{3})=0$
are plotted. Depending on the choice of $k$, the resulting curves
are either parabolas, hyperbolics, or collapse to straight lines on
the particular plane. However, all of the curves intersect at the
solution point as expected.

\section{Discussion}

The main results of the previous sections can be summarized as follows:
Using the combination matrix $C^{\mathbf{pq},\mathbf{k}}$, the Navier-Stokes
equations can be rewritten in terms of intersecting quadratic polynomials
(\ref{eq:NS_Conic1}), where the solutions to the system are determined
by the number of intersections to the polynomial system. This is shown
both for the case of homogeneous isotropic turbulence and planar channel
flow using the Chebyshev polynomial basis. Once the polynomial coefficients
are known, a solvability condition can be derived, and the solution
to these systems can then be solved using standard methods from commutative
algebra. For instance, Buchberger's algorithm can be applied to transform
the polynomial system into a Groebner basis, which can then be solved
to reveal the common roots. This process is demonstrated for a simple
nonlinear differential equation.

Several key points are worth noting regarding this analysis. While
we show that the combination can be defined for the Fourier and Chebyshev
bases, it can also be extended to other orthogonal polynomials by
considering the Clebsch-Gordan or standard linearization problem \citep{Ruiz_linearizationformula}.
Secondly, the method used to determine the Groebner basis in section
\ref{sec:Groebner-Basis}, while useful to illustrate the procedure
on simple problems, does not efficiently scale to larger problems.
Buchberger's original algorithm is numerically unstable and diffcult
to parallelize to tackle large systems of polynomials. While more
efficient methods of finding a Groebner basis are available, it is
also possible to use multivariate resultants (similar to those discussed
in section \ref{sec:Solvability}) to solve the polynomial system
(\ref{eq:NS_Conic1}). These methods \citep{hanzon2006proceedings}
convert the system to a standard eigenvalue problem, wihch can then
be solved using standard approaches from numerical linear algebra
to yield all possible solutions to the original problem. Implementing
such an algorithm is more involved than the method described in Appendix
\ref{sec:Appendix}, but if done properly, will yield the same result
for the Navier-Stokes.

\appendix

\section{\label{sec:Appendix}Appendix: Buchberger's Algorithm}

As mentioned in section \ref{sec:Groebner-Basis}, Buchberger's algorithm
was the first known method for transforming a set of polynomials into
a Groebner basis given a particular monomial ordering. This algorithm
relies on the creation of an S-polynomial for two polynomials $f$
and $g$, defined as

\[
S(f,g)=\textrm{LCM}(\textrm{LPP}(f),\textrm{LPP}(g))\left(\frac{f}{\textrm{LM}(f)}-\frac{g}{\textrm{LM}(g)}\right)
\]
where $\textrm{LPP}(f)$ is the leading power product of the polynomial
$f$, $\textrm{\textrm{LM}}(f)$ is the leading monomial of $f$,
and $\textrm{LCM}(f,g)$ is the least common multiple of $f$ and
$g$. From this, Buchberger's method for generating a Groebner basis
can be roughly implemented as described in algorithm \ref{alg:Buchberger's-algorithm}.

\begin{algorithm}[h]
Input: $F=(f_{1},...,f_{s})$ 

Output: a Groebner basis $G=\left\{ g_{1},...,g_{t}\right\} $ for
$I=\left\langle F\right\rangle $, with $F\subset G$ 

$\quad$LET G := F 

$\quad$REPEAT 

$\quad$$\quad$$G'$ := $G$ 

$\quad$$\quad$FOR each $\left(p,q\right)$ pair $p\neq q$ in $G'$
DO 

$\quad$$\quad$$\quad$S := REM\{$S(p,q)$\} 

$\quad$$\quad$$\quad$IF $S\neq0$ THEN $G=G\cup\left\{ S\right\} $ 

$\quad$UNTIL $G=G'$

\caption{\label{alg:Buchberger's-algorithm}Buchberger's algorithm, as described
in \citet{cox2006using} }
\end{algorithm}

\bibliographystyle{jfm}
\bibliography{references}

\end{document}